\documentclass[aps, pra,
reprint,
superscriptaddress,
twocolumn,
floatfix]{revtex4-2}
\usepackage[T1]{fontenc}
\usepackage{amsfonts}
\usepackage{CJKutf8}
\usepackage{amssymb}
\usepackage[toc,page]{appendix}
\usepackage{amsmath,mathtools,amsthm,amssymb}
\usepackage{bm}
\usepackage{adjustbox}
 
\usepackage{tabularx}
\usepackage{units}
\usepackage{complexity}
\usepackage{graphicx}
\usepackage{color}
\usepackage{xcolor}
\usepackage{bbold}
\usepackage{complexity}
\usepackage{booktabs}
 
\usepackage[linesnumbered,ruled,vlined]{algorithm2e}
\usepackage{listings}

\usepackage{lipsum} 
\usepackage{enumitem}
\usepackage{comment}
\usepackage{pifont}
\usepackage{times}
\usepackage{comment}
\usepackage{array}
\newcolumntype{H}{>{\setbox0=\hbox\bgroup}c<{\egroup}@{}}
\usepackage{graphics}
\usepackage[normalem]{ulem}

\usepackage{complexity}
\newcommand{\mc}[1]{\mathcal{#1}}

\definecolor{myrefcolor}{RGB}{242, 10, 10}
\definecolor{myurlcolor}{RGB}{255, 138, 48}

\newcommand{\kx}{\kappa_{\text{\scriptsize ex}}}
\newcommand{\ki}{\kappa_{\text{\scriptsize i}}}

\usepackage[toc,page]{appendix}

\usepackage{braket}
\usepackage{physics}
\usepackage{csquotes}
\usepackage{thmtools, thm-restate}

\newcommand{\bigb}[1]{\big( #1\big)}

\newcommand{\aout}[1]{a_{\text{out}, #1}}
\newcommand{\baout}[1]{\bar a_{\text{out}, #1}}

\usepackage{longtable}
 
\usepackage{url}
 
\usepackage[colorlinks=true, linkcolor=blue, citecolor=blue, urlcolor=blue]{hyperref}

\begin{document}
\preprint{ }
\title{Fock-state preparation based on amplitude amplification in cavity QED}

\author{Sharoon Austin}
\affiliation{Joint Quantum Institute, NIST/University of Maryland, College Park, MD, 20742, USA}\affiliation{Joint Center for Quantum Information and Computer Science, NIST/University of Maryland, College Park, MD, 20742, USA}
\author{Zhi-Yuan Wei (\begin{CJK*}{UTF8}{gbsn}魏志远\end{CJK*})}
\affiliation{Joint Quantum Institute, NIST/University of Maryland, College Park, MD, 20742, USA}\affiliation{Joint Center for Quantum Information and Computer Science, NIST/University of Maryland, College Park, MD, 20742, USA}
\author{Kartik Srinivasan}
\affiliation{Joint Quantum Institute, NIST/University of Maryland, College Park, MD, 20742, USA}
\affiliation{Microsystems and Nanotechnology Division, Physical Measurement Laboratory, National Institute of Standards and Technology,
Gaithersburg, Maryland 20899, USA}
\author{Alexey V. Gorshkov}
\affiliation{Joint Quantum Institute, NIST/University of Maryland, College Park, MD, 20742, USA}\affiliation{Joint Center for Quantum Information and Computer Science, NIST/University of Maryland, College Park, MD, 20742, USA}

\begin{abstract}
In this work, we develop a coherent control technique for cavity QED based on amplitude amplification. 
We consider two physical platforms. In the first setting, we study a three-level quantum emitter coupled to a single mode of an optical cavity and introduce a protocol for producing traveling single photons based on oblivious amplitude amplification. As the key ingredient of our protocol, we propose an extension of oblivious amplitude amplification which uses reflection unitaries solely on the signal qubit, along with $U$ and $U^\dagger$, where the unitary $U$ prepares the initial state. Our approach improves the scaling of the single-photon-generation protocol length from $N\sim 1/p$ to $N\sim 1/\sqrt{p}$, with $p$ denoting the success probability of obtaining a short single photon from a single application of the weak control pulse. Furthermore, our protocol also reduces the error from intrinsic cavity loss compared to protocols using a single strong control pulse in various experimentally relevant regimes, suggesting the application of our methods for error reduction. In the second setting, we consider a superconducting qubit coupled to a single bosonic mode of a microwave cavity in a circuit QED architecture in the dispersive regime for preparing Fock states. Using fixed-point amplitude amplification, we obtain a protocol for preparing Fock states whose length scales as $O(n^{1/4})$, where $n$ is the number of photons. Additionally, as an application of our methods for state preparation, we describe a protocol for preparing NOON states.
\end{abstract}

\volumeyear{ }
\volumenumber{ }
\issuenumber{ }
\eid{ }
\date{\today}
\startpage{1}
\endpage{10}
\maketitle

\section{Introduction}

Achieving high-efficiency single-photon sources is essential for several key quantum information processing tasks, including quantum metrology \cite{PhysRevLett.121.060506,PhysRevA.54.R4649, Pirandola_2018}, quantum communication \cite{QRepeat, RevModPhys.74.145, pittaluga2024}, and quantum computation \cite{pan_multiphoton_2012, kok_linear_2007, PhysRevLett.92.127902, PsiQ}. Crucially, it is important to produce single photons with sufficiently high fidelities to perform such tasks with a quantum advantage \cite{PhysRevLett.114.170802, Bhaskar, Sarafusion}.

Preparing Fock states of cavity modes is an important primitive in circuit quantum electrodynamics (circuit QED) \cite{HofheinzFock, WeakFock}. Moreover, both large-photon-number Fock states \cite{Wolf_2019,Deng2024} and large-photon-number superpositions of multimode Fock states, such as NOON states \cite{NOONseth}, can be used for quantum sensing beyond the standard quantum limit.

Single photons can be produced via coherent control techniques \cite{axelnet, alexeyP1, PhysRevB.97.125303} in quantum dots \cite{sweeney2014cavity} and atoms \cite{PhysRevA.87.063805} coupled to high-finesse cavities with auxiliary classical fields. Moreover, protocols have been developed that use classical fields to control the polarization \cite{polalt} and temporal shape \cite{waveform, qdwave} of the emitted single photons. Generating relatively short single photons is crucial, as it enables photonic gates to operate at the high clock rates required for practical quantum information processing. 
While temporal compression of single photons after their generation is possible, it remains experimentally challenging and incurs substantial overhead~\cite{reshaping1,2hq3-t534}. On the other hand, if a three-level system is coupled to a cavity and one of the transitions is driven by a classical laser, the duration of the emitted single photon can be much shorter than the spontaneous emission lifetime of the system. However, generating such short single photons requires strong classical control pulses, with the required control-pulse Rabi frequency scaling as $\sim g/\sqrt{ T}$, where $g$ is the cavity coupling strength to the emitter and $T$ is the single-photon duration~\cite{vqed}.
Thus, to keep the pulse duration relatively short, a large control pulse power is required. In this work, we focus on the case where we choose smaller values of $T$ but do not have access to sufficiently large values of $\Omega_0$.
We also note that the ability to implement a single-photon source without large classical Rabi frequencies reduces various experimental constraints.
For example, using smaller classical Rabi frequencies makes it possible to choose a three-level system in the emitter without concern for off-resonant transitions or multilevel effects. This also enables the choice of atomic transitions with small dipole moments, allows transverse application of the control beam without coupling it into the cavity mode, and reduces heating and charge noise in quantum dots.

In circuit QED, in the dispersive limit \cite{PhysRevA.69.062320, circuitQED,RevModPhys.93.025005, fong2025engineering}, selective number-dependent arbitrary phase (SNAP) gates and cavity displacements can be used in unitary sequences to produce Fock states \cite{PhysRevLett.115.137002, PhysRevA.92.040303}. Recent work on reducing the length of such unitary sequences has used numerical optimization methods \cite{PhysRevA.92.040303, HeeresGrape,PRXQuantum.3.030301, fösel2020efficientcavitycontrolsnap}. Previous works have used resonant transmon–cavity interactions \cite{FastSimpleNoon} and three-level transmons \cite{FastSimpleNoon, PhysRevA.107.042412, Merkel_2010} to generate NOON states.

\begin{figure}[h]
		\includegraphics[
		width=0.9\columnwidth
		]{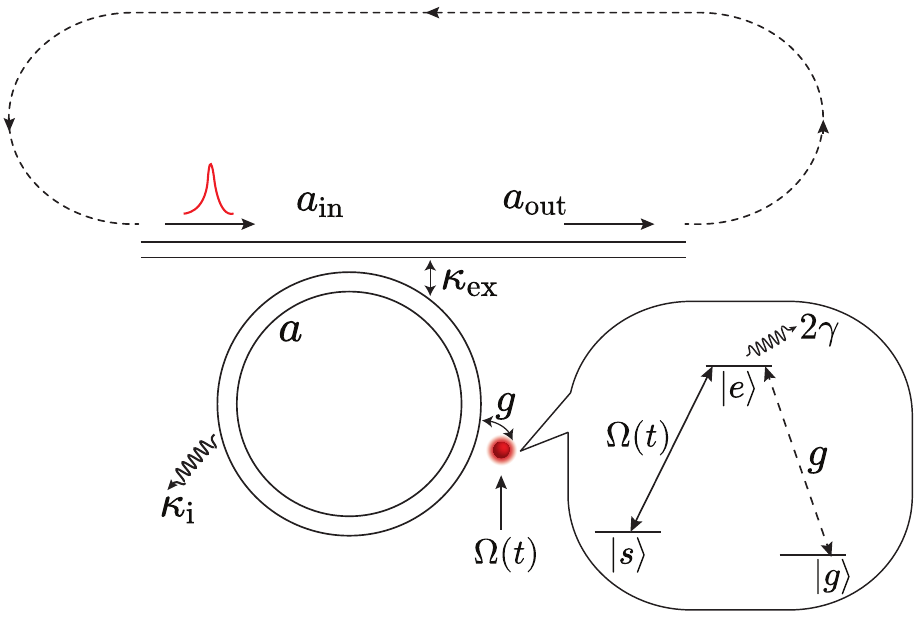}
	\caption{The atom is coupled to the cavity mode $a$, the cavity mode is coupled to the waveguide with coupling $\kx$, and the cavity has an intrinsic loss rate $\ki$.  As shown on the right, the atomic levels $\ket{s}$ and $\ket{e}$ are coupled resonantly by a classical laser pulse (shown by a solid-black line) with Rabi frequency $\Omega(t)$, while the atomic levels $\ket{e}$ and $\ket{g}$ are coupled resonantly by the cavity with single-photon Rabi frequency $g$ (shown by a dashed-black line). The state $\ket{e}$ decays with rate $2\gamma$. Each step of the amplification protocol is implemented by sending the single-photon output back to the cavity while applying different versions of the control pulse $\Omega(t)$ from Eq.~(\ref{G3control}).}
	\label{atomiclevel}
\end{figure}

In this work, we use Grover amplification for state preparation. Previous work \cite{3fzf-wsr2} has shown that experimental protocols based on Grover amplification can be used to prepare useful quantum states of light. In this work, we propose an extension of oblivious amplitude amplification \cite{Berry_2014}, which we call \textit{very oblivious amplitude amplification} and which uses reflection unitaries acting solely on the signal qubit \cite{Berry_2014}, along with applications of $U$ and $U^\dagger$, where $U$ is the unitary that prepares the initial state.  Building on previous methods \cite{3fzf-wsr2} and using very oblivious amplitude amplification, we  design a protocol for single-photon generation in a $\Lambda$-type system coupled to a cavity.  The protocol uses weak control pulses combined with multiple passes of the generated single-photon output through the cavity to coherently amplify the single-photon amplitude in the desired mode. Our protocol can be implemented in any physical platform containing a three-level quantum emitter, as shown in Fig.~\ref{atomiclevel}, in which one branch is coherently driven by a classical field and the other branch is coupled to a cavity, e.g., quantum dots with high-coherence optical transitions \cite{Najer2019} and superconducting three-level systems coupled to resonators \cite{PhysRevX.4.041010}. We improve upon the following probabilistic protocol for generating a single photon: apply a weak control pulse, perform an atomic-state measurement that heralds single-photon generation, and repeat the process until a single photon is obtained. Our $k$-step unitary protocol for generating relatively short single photons uses control pulses with Rabi frequencies that scale inversely with $k$.
In contrast to previous approaches that employ a single strong control pulse for single-photon generation \cite{alexeyP1, made}, our protocol exhibits a qualitatively different dependence of the single-photon loss on system parameters, which is advantageous in many experimentally relevant scenarios \cite{vqed}. Additionally, for a superconducting qubit dispersively coupled to a bosonic mode, we use the fixed-point version of amplitude amplification \cite{Yoder_2014} to derive an explicit analytic unitary sequence consisting of SNAP gates and cavity displacements to produce Fock states. Our protocol duration scales as $O(n^{1/4})$, where $n$ is the number of photons, and therefore achieves a quadratic improvement compared to previous work, which achieves $O(n^{1/2})$ scaling \cite{PhysRevA.92.040303}. For two superconducting qubits with strong dispersive coupling to their respective cavity modes, we show that our methods can be extended by employing the beam-splitter interaction between cavity modes \cite{bs0, bs1, bs2, PRXQuantum.4.020355} and using a cat state as the initial state \cite{catstate1, catstate2} to produce NOON states \cite{PhysRevA.65.052104, PhysRevA.107.042412}. Our protocol is simpler compared to previous approaches \cite{FastSimpleNoon, PhysRevA.107.042412, Merkel_2010} because it does not require resonant transmon–cavity interactions \cite{PhysRevLett.115.137002} or a three-level structure for the transmons.

This paper is organized as follows. In Sec.~\ref{methods}, we present the theoretical framework and results covering amplitude amplification. In Sec.~\ref{ress}, we describe the details of the two physical platforms and present the main results. In Sec.~\ref{conout}, we discuss the outlook and conclusion. Finally, in the Appendices, we present details omitted from the main text.
\section{Methods}\label{methods}

In this section, we first review key results for two variants of amplitude amplification \cite{nielsen2010, Yoder_2014}. We then present a simplified version of oblivious amplitude amplification \cite{Berry_2014}. Finally, we propose an extension of oblivious amplitude amplification tailored to our physical implementation.

We first describe the amplitude amplification algorithm \cite{GroverPrl}. We consider the orthogonal states $\{\ket{\phi_1}, \ket{\phi_0} \}$ and the superposition $\ket{\psi_0} = \sin\theta \ket{\phi_1} + \cos \theta \ket{\phi_0}$, where $\theta \ll \pi/2$. To amplify the amplitude of the state $\ket{\phi_1}$, we implement the Grover operator defined as follows \cite{nielsen2010}:
\begin{align}
G_0 &= R_iR_1, \label{grover}
\end{align}
where
\begin{align}
R_1 &= I - 2\ketbra{\phi_1}{\phi_1}, \label{Rg0}\\
R_i &=   I - 2\ketbra{\psi_0}{\psi_0}   \label{Ridef}.
\end{align}
Here $R_1$ and $R_i$ are reflections about the states $\ket{\phi_1}$ and $\ket{\psi_0}$, respectively. We can then write $G_0$ in the basis $\{\ket{\phi_1}, \ket{\phi_0} \}$ as follows:
\begin{align}
G_0&= -\begin{pmatrix} \cos 2\theta & \sin 2 \theta\\
- \sin 2 \theta & \cos 2 \theta
\end{pmatrix}, \label{G0}
\end{align}
so that applying $U_{G_0}=(-1)^k G_0^k$ to the initial state $\ket{\psi_0}$
amplifies the amplitude of the state $\ket{\phi_1}$ as follows:
\begin{align}
U_{G_0} \ket{\psi_0}= \sin[(2k+1)\theta] \ket{\phi_1} + \cos[(2k+1)\theta] \ket{\phi_0}. \label{Ug0}
\end{align}
Moreover, the choice
\begin{align}
\theta = \frac{\pi}{2(2k+1)} \label{thetak}
\end{align}
gives us the state $\ket{\phi_1}$ after $k$ iterations of the unitary $G_0$. We remark here that for the applications considered in this work, preparing an initial state with a fine-tuned value of $\theta$ to achieve perfect state preparation might not be straightforward.

In amplitude amplification as described above, the success probability of the protocol depends on the precise value of the overlap $\braket{ \phi_1}{\psi_0}$ through Eq.~(\ref{Ug0}).
We now describe fixed-point amplitude amplification \cite{Yoder_2014}, in which the probability of obtaining the `good state' is insensitive to the precise value of $\theta$. Here the $l$-step amplification unitary $U_{G_1}$ is given by 
\begin{align}
 U_{G_1}
&= (-1)^l \prod_{j=1}^{l}R_i(\alpha_j )R_1(\beta_j), \label{G2fp}
\end{align}
where the unitaries are ordered from right to left.
Here, $R_1(\beta)$, $R_i(\alpha)$ are generalized reflections about the state $\ket{\phi_1}$ and the initial state $\ket{\psi_0}$, respectively, defined as follows:
\begin{align}
R_1(\beta) &= I - (1-e^{i\beta })\ketbra{\phi_1}{\phi_1}, \label{fpgood}\\
R_i(\alpha) &= I - (1-e^{i\alpha })\ketbra{\psi_0}{\psi_0}. \label{fpinitial}
\end{align}
For a given $\delta$, there exist phases $\alpha_j$, $\beta_j$ (with exact analytic expressions) such that success probability $\abs{\bra{\phi_1}U_{G_1}\ket{\psi_0}}^2 \geq 1-\delta^2$ with \cite{Yoder_2014}
\begin{align}
 l  \geq [\log(2/\delta)(\sin\theta)^{-1}-1]/2  \label{lbound}.
\end{align}

We now consider oblivious amplitude amplification, used in various applications including Hamiltonian simulation \cite{Berry_2014, PhysRevLett.114.090502} and quantum algorithms \cite{vasconcelos2025}. We consider a simplified setting with two registers: the first contains the signal qubit, and the second contains an arbitrary quantum state. Let $\ket{\phi_1}$ denote the ``good'' state that we wish to amplify, and let $\ket{\phi_0^\prime}$ denote an arbitrary state that is not necessarily orthogonal to $\ket{\phi_1}$. The initial state $\ket{\bar{\psi}_0}$ is given by
\begin{align}
|\bar \psi_0 \rangle = \bar U \ket{0}\ket{\phi_0} = \sin\theta \ket{1}\ket{\phi_1} + \cos \theta \ket{0}\ket{\phi_0^\prime}.\label{oaasetup}
\end{align}
By using the structure of the unitary $\bar U$ in that the state $\ket{\phi_1}$ always comes with the signal qubit state $\ket{1}$, we can replace the reflection about the good state in Eq.~(\ref{Rg0}) with a reflection about the signal qubit state instead. Using
\begin{align}
\bar R_1 &= I - 2\ketbra{0}{0}, \label{ronebar}\\
R_i &= I-2|\bar \psi_0 \rangle \langle \bar \psi_0|,
\end{align}
gives us the unitary $G_2 =  R_i \bar R_1$, which can be written in the basis $\{ \ket{1}\ket{\phi_1}, \ket{0}\ket{\phi_0^\prime}\}$ as
\begin{align}
G_2&= \begin{pmatrix} \cos 2\theta & \sin 2 \theta\\
- \sin 2 \theta & \cos 2 \theta
\end{pmatrix}. \label{G1}
\end{align}
$G_2$ has the same form as $G_0$ in Eq.~(\ref{G0}) up to a minus sign. Here, $G_2$ contains $\bar R_1$, the reflection about the signal-qubit state, instead of $R_1$ in $G_0$.

\subsection{Very oblivious amplitude amplification}\label{voaasub}

In this subsection, we propose an extension of oblivious amplitude amplification that simplifies $R_i$ and that we call very oblivious amplitude amplification. We first note that $G_2$ still contains $R_i$, a reflection about the initial state $|\bar \psi_0 \rangle$ that involves both registers. To simplify this reflection, we introduce additional physically motivated structure that allows $R_i$ to be replaced by a simpler unitary.
We assume the initial state 
\begin{align}
|\bar \psi_0 \rangle = \bar U \ket{0}\ket{\phi_0} = \sin\theta \ket{1}\ket{\phi_1} + \cos \theta \ket{0}\ket{\phi_0},\label{oaasetup2}
\end{align}
with the conditions $\braket{\phi_0}{\phi_1}=0$ and $[\bar U, N]=0$, where
\begin{align}
&N = \ketbra{0}{0}\otimes I + I\otimes(I-\ketbra{\phi_0}{\phi_0}).
\end{align}
Here, the condition $[\bar U, N] = 0$ corresponds to a conserved observable $N$, which constrains the dynamics under $\bar U$ to a specific subspace.
We can then simplify the unitary $G_2 = R_i \bar R_1$ from Eq.~(\ref{G1}) by using the structure of $\bar U$ to simplify $R_i$ as follows.
We first write $G_2 |\bar \psi_0 \rangle$ as follows:
\begin{align}
 G_2|\bar \psi_0 \rangle&=R_i \bar R_1 |\bar \psi_0 \rangle \\&=  \bar U(I - 2\ketbra{0}{0}\otimes \ketbra{\phi_0}{\phi_0})\bar U^\dagger \bar R_1|\bar \psi_0 \rangle  \label{g2eqq}.
\end{align}
Here, $\bar U^\dagger \bar R_1 |\bar \psi_0 \rangle$ lies in the eigenspace of $N$ with eigenvalue 1, i.e., the span of $\{\ket{0}\ket{\phi_0}, \ket{1}\ket{\phi_0^{\perp}} \}$, where $\ket{\phi_0^\perp}$ is any state orthogonal to $\ket{\phi_0}$. This follows from the facts that $[\bar U^\dagger, N]=[\bar R_1, N]=0$, and $|\bar \psi_0 \rangle$ is an eigenstate of $N$ with eigenvalue 1, implying that $\bar U^\dagger \bar R_1 |\bar \psi_0 \rangle$ is also an eigenstate of $N$ with eigenvalue 1. Since $ \bar U^\dagger \bar R_1 \lvert \bar \psi_0\rangle $ lies in the span of $ \{\lvert 0\rangle \lvert \phi_0\rangle, \lvert 1\rangle \lvert \phi_0^\perp\rangle \}$, the reflection about the state $ \lvert 0\rangle \lvert \phi_0\rangle $ in Eq.~(\ref{g2eqq}) can be replaced by a reflection about the state $ \lvert 0\rangle$. Therefore, the reflection operator \( R_i \) can be replaced by $ \bar U \bar R_1 \bar U^\dagger $. Moreover, the resulting state  $G_2|\bar \psi_0 \rangle$ is also an eigenstate of $N$ with eigenvalue 1. Therefore, we can replace $G_2$ with
\begin{align}
G_3 = \bar U \bar R_1 \bar U^\dagger \bar R_1 \label{gthreee},
\end{align}
which implements what we call very oblivious amplitude amplification,
using signal-qubit reflections along with applications of $\bar U$ and $\bar U^\dagger$.

\section{Results}\label{ress}

In this section, we present two protocols implemented in different physical platforms. In Sec.~\ref{cqed}, we study the generation of traveling single photons in a $\Lambda$-type cavity QED system using amplitude amplification. In Sec.~\ref{res1}, we turn to superconducting circuit QED, where we describe protocols for generating Fock states and NOON states using fixed-point amplitude amplification.

\subsection{Single-photon generation in a $\Lambda$-type cavity QED system}\label{cqed}
In this subsection, we describe how to prepare a traveling single-photon state with an emitter-cavity system using amplitude amplification. We first describe the emitter-cavity system.
As shown in Fig.~\ref{atomiclevel}, we consider a $\Lambda$-type system coupled to a cavity. The states $\ket{s}$ and $\ket{e}$ are coupled resonantly by a classical laser with real Rabi frequency $\Omega(t)$, and the states $\ket{e}$ and $\ket{g}$ are coupled resonantly by the cavity mode with single-photon Rabi frequency $g$. The cavity mode has annihilation operator $a$. The cavity is coupled to the waveguide with rate $\kx$, the intrinsic cavity loss rate is $\ki$, and the spontaneous decay rate of the excited state $|e\rangle$ is $2\gamma$. In a rotating frame, the system is described by the following no-jump effective non-Hermitian Hamiltonian $H$ and input-output relation \cite{kuzmich}:
\begin{align}
H
&= (ga \ketbra{e}{g}+\Omega(t) \ketbra{e}{s} + \text{H.c.}) -i\gamma \ketbra{e}{e}, \label{dyn1}\\
\dot{a} &= -i[a, H]-\kappa a -\sqrt{2\kx}a_{\text{in}}(t), \label{dyn2}\\
a_{\text{out}}(t) &= a_{\text{in}}(t) + \sqrt{2\kappa_{\text{ex}}} a(t), \label{dyn3}
\end{align}
where H.c.~denotes the Hermitian conjugate, $\kappa = \ki+\kx$ with $\ki < \kx$. As discussed at the end of this subsection, we also consider the case where the waveguide is lossy such that the transmission probability per trip through the waveguide is $1-\epsilon_t$, with $\epsilon_t>0$. Here, $a_{\text{in}}(t)$ and $a_{\text{out}}(t)$ are the input and output modes, respectively. As discussed in Appendix \ref{ufder}, since our dynamics is constrained to the single-excitation subspace, $a_{\text{in}}(t)$ and $a_{\text{out}}(t)$ appear as complex numbers in our analysis \cite{alexeyP1}. We remark here that any unknown relative phase between the cavity and classical laser fields can be absorbed into a redefinition of the atomic states $\ket{e}$ and $\ket{s}$. We define the efficiency of single-photon retrieval $\eta_2$ as $\eta_2 = \max_\tau  \lvert \int_{-\infty}^\infty a_{\text{out}}(t) h^*(t+\tau) dt \rvert^2$, where $a_{\text{out}}(t)$ is the single-photon output of our protocol. Here, $\eta_2$ measures the efficiency of single-photon retrieval in the desired mode $h(t)$. As discussed in detail later in this section, this definition accounts for the fact that our single-photon output can have small time translations relative to the desired mode $h(t)$ due to non-adiabatic errors. Using results in Refs.~\cite{made, gauss, vqed}, we can design the control pulse $\Omega(t)$ to retrieve a single photon with mode function $h(t)$ and chosen efficiency $\eta <\eta_{\text{max}}$ [i.e., $a_{\text{out}}(t) = \sqrt{\eta}h(t)$], where
\begin{align}
\eta_{\text{max}} =\frac{\kx}{\kappa}\Big(\frac{C}{1+C} \Big), \label{effc}
\end{align}
with $C=g^2/(\kappa \gamma)$. Here, the efficiency $\eta_{\text{max}}$ is achieved in the limit $\kx T \gg 1$. Values of $\eta<1$ correspond, physically, to either amplitude left in the state $\ket{s, 0}$ or loss through $\ki$ and $\gamma$. We note that choosing larger values of $\kx$ reduces the required control pulse amplitude, which scales as $g/\sqrt{\kx T}$. However, it also lowers the cooperativity $C$ and, consequently, reduces the maximum achievable efficiency given by Eq.~(\ref{effc}).

The required control pulse magnitude $|\Omega(t)|$ for $\eta \approx \eta_{\text{max}}$ scales as $g/\sqrt{\kx T}$. We consider the case where $T$ is small and sufficiently large values of $|\Omega(t)|$ are inaccessible. We first consider the loss-free case in which $\ki = \gamma=0$. For a given $\eta \ll 1$, one can choose a weak control pulse that implements the unitary $\bar{U}$ and prepares the atom–photon state \cite{made,gauss,vqed}
\begin{align}
\ket{\bar{\psi}_0} = \bar U\ket{s, 0_p} = \sin \theta \ket{g, 1_h} + \cos \theta \ket{s, 0_p} \label{apstate},
\end{align}
where $\sin \theta=\sqrt{\eta}$, $\ket{0_p}$ denotes the bosonic vacuum state, and $\sin\theta \ket{ 1_h}$ denotes the single-photon state in the waveguide with $a_{\text{out}}(t) = \sqrt{\eta} h(t)$. Using results from Refs.~\cite{made, vqed}, we choose the control pulse implementing $\bar U$ for an arbitrary $h(t)$ such that $\theta$ satisfies Eq.~(\ref{thetak}) for a given $k$. As discussed later, the value of $k$ can be chosen to minimize the size of the required $\Omega(t)$ or to minimize the single-photon loss due to $\ki$ and $\gamma$.
We first note that we can identify $\{\ket{0}, \ket{1} \} \leftrightarrow \{\ket{s}, \ket{g} \}$ and $\{\ket{\phi_0}, \ket{\phi_1} \} \leftrightarrow \{\ket{0_p}, \ket{1_h} \}$, corresponding to Eq.~(\ref{oaasetup}). Here, $\ket{1_h}$ is the state we want to prepare.
The unitary $\bar U^\dagger$ can be implemented using the composition $\bar U^\dagger = \mathcal{T}U_b \mathcal{T}$, where $U_b$ is effected by applying a modified control pulse in the Hamiltonian in Eq.~(\ref{dyn1}) \cite{alexeyP1}. Here $\mathcal{T}$ is defined as the operator that effects the transformation $f(t) \rightarrow f(T-t)$ on the single-photon mode, assumed to be real.  Details of the derivation of $\bar U^\dagger$ and $U_b$ are presented in Appendix \ref{ufder}. Moreover, since $\bar{U}^\dagger$ is implemented via the dynamics in Eqs.~(\ref{dyn1})--(\ref{dyn3}), it conserves the number of excitations $N$ in the atom–waveguide system and thus satisfies $[\bar{U}^\dagger, N] = 0$, where
\begin{align}
N = \ketbra{s}{s}\otimes I + I \otimes(I - \ketbra{0_p}{0_p}).
\end{align}
Here, $N$ corresponds to the excitations before and after the atom-cavity interaction.
We also note that $\braket{0_p}{1_h}=0$. We can then employ \textit{very oblivious amplitude amplification}, as described in Sec.~{\ref{voaasub}}, to obtain the Grover unitary
\begin{align}
G_3 &=\bar U \bar R_1 \bar U^\dagger \bar R_1\\
&= \bar U \bar R_1 \mc{T}U_b \mc{T} \bar R_1\\
&=\bar U \mathcal{T}W\mathcal{T}, \label{G3control}
\end{align}
where $W = \bar R_1 U_b \bar R_1$ corresponds to dynamics implemented in Eqs.~(\ref{dyn1})--(\ref{dyn3}) using the control pulse $\Omega_W(t)=\Omega(T-t)$. For more details of the derivation of $\Omega_{W}(t)$ used to implement $W$, see Appendix \ref{ufder}. The Grover sequence composed of $k$ applications of $G_3$ [Eq.~(\ref{gthreee})] can be written as
\begin{align}
U_{G_3}  = S_{2k}\hdots S_2S_1S_0, \label{ug3}
\end{align}
which is applied to the initial state $\ket{s, 0_p}$. Here, $S_j = \mathcal{T}W\mathcal{T}$ for odd $j$, and $S_j=\bar U$ for even $j$. As shown in Appendix~\ref{appbee}, for $k \gg 1$, the control pulse magnitude $|\Omega(t)|$ is proportional to $\sin\theta$ [Eq.~(\ref{thetak})] and therefore vanishes as $1/k$.

\begin{figure}[ht]
		\includegraphics[
		width=\columnwidth
		]{ 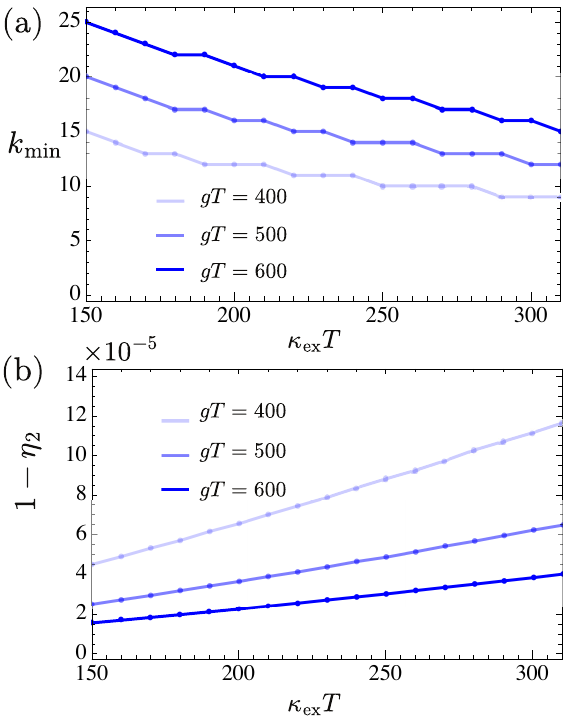} 
    \caption{In (a), for our protocol, we show $k_{\text{min}}$, the value of $k$ that minimizes the inefficiency $1-\eta_2$, where $\eta_2 = \max_\tau \lvert \int_{-\infty}^\infty a_{\text{out}}(t) h^*(t+\tau) dt \rvert^2$, versus $\kx T$ for various values of $g$ in the lossless case ($\ki = \gamma =0$). Here, the $\mc{T}$ operations are omitted. The value of $T$ is kept fixed. Here, $\eta_2$ is obtained from numerical solutions of Eqs.~(\ref{dyn1})--(\ref{dyn3}) with the single-photon mode $h(t) =\sqrt{8/(3T)}\sin^2(\pi t/T)$. As $[\kx/g, 1/(gT)]$ decrease, the value of $k$ that minimizes $1-\eta_2$ increases. In (b), we show the value of $1-\eta_2$ obtained at $k=k_{\text{min}}$ for the corresponding values of $(gT, \kx T)$ shown in (a).} 
	\label{nonAd}
\end{figure}

We now consider the case where the $\mathcal{T}$ operation cannot be implemented on the single-photon mode, as it requires significant experimental overhead, such as nonlinear optics \cite{Lavoie, 10.1117/12.2582621}. In the limit $k \gg 1$, where $k$ is the length of the Grover sequence, the single-photon output at each step $S_j$ [Eq.~(\ref{ug3})], where $j = 0, 1, \hdots, 2k$, can be written as [cf.~Appendix \ref{appbee}]
\begin{align}
a_{\text{out}, j} (t) \approx h(t) \sin \theta \sum_{m=0}^{j}\cos(m\theta),
\label{aoutj}
\end{align}
where $\kx/(g^2T) \ll 1$, and $\theta$ satisfies Eq.~(\ref{thetak}). This implies that, at every step of the protocol, the single-photon output is invariant under $\mathcal{T}$ to high accuracy if we choose $h(t) = h(T-t)$.
As a result, the $\mathcal{T}$ steps can be omitted at the expense of small non-adiabatic errors due to finite $T$, which scale as $\kx/ (g^2T)$, in the final single-photon output [see Appendix \ref{appbee} for more details].
Moreover, numerics, as shown in Fig.~\ref{nonAd}, indicate that there is a range of practical values of $k$, i.e., values of $k$ that satisfy $k \gg1$ while remaining sufficiently small such that the inefficiency $1-\eta_2$ of our protocol is small. 
As shown in Fig.~\ref{nonAd}, as $[\kx/g, 1/(gT)]$ decrease, the value of $k$ that minimizes the inefficiency $1-\eta_2$ increases, i.e., the non-adiabatic losses start accumulating at larger values of $k$.
The inefficiency $1-\eta_2$ has an upper bound of $O(k^2\kx^2/(g^4 T^2))$ [see Appendix \ref{appbee} for more details]. 

We now evaluate our single-photon amplification scheme from above in the presence of losses $\gamma$ and $\ki$.
As shown in Appendix \ref{LossDerivation}, in the limits $k \gg 1$ and $\kx/(g^2 T) \ll 1$, where $k$ is the length of our Grover sequence, the single-photon loss to first order in the loss rates is
\begin{align}
 1-\eta_3  &= \frac{\kappa_{\text{i}}}{\kx} (k+1) \sin^2\theta + 4kC^{-1}_{\text{ex}}, \label{plossEq}
\end{align}
where $\eta_3 = \int_{-\infty}^\infty \abs{a_{\text{out}}(t)}^2dt $, with $a_{\text{out}}(t)$ the single-photon output at the end of the protocol in Eq.~(\ref{ug3}). The first term corresponds to intrinsic cavity loss, and the second term corresponds to loss due to the excited state decay, with $\theta = \pi/[2(2k+1)]$ [from Eq.~(\ref{thetak})] and $C_{\text{ex}}= g^2/(\kx \gamma)$. Remarkably, in the case where $\gamma = 0$, the single-photon loss due to $\ki$ vanishes as $\sim 1/k$. This fact can be explained as follows. As shown in Appendix \ref{appbee}, in the lossless case, the cavity amplitude in the single-photon state $\ket{g, 1}$ during the $j$th step of the sequence $U_{G_3}$, denoted as $c_{gj}(t)$, can be written as
\begin{align}
c_{gj}(t)&\approx  h(t)\sin\theta \cos(j\theta)/\sqrt{2\kx},\>\>\> j=0, 1, \hdots, 2k, \label{cgjexp}
\end{align}
which means that the cavity amplitude decreases as the protocol progresses. Since the single-photon loss through the cavity during the $j$th step scales as $\sim \ki\sin^2\theta \cos^2(j\theta)$, the total loss scales as $ \sim \ki \sin^2 \theta \sum_{j=0}^{2k}\cos^2(j\theta)$ which goes to zero as $\ki  (k+1)^{-1}$. We now consider the case where $\gamma$ is nonzero. In this case, increasing $k$ leads to an effective cooperativity $C_{\text{ex}}/[4k]$ which decreases with $k$. Here, the decrease in cooperativity reflects the fact that multiple atom–cavity interactions increase the single-photon loss via $\gamma$.

\begin{figure}[t]
		\includegraphics[
		width=\columnwidth
		]{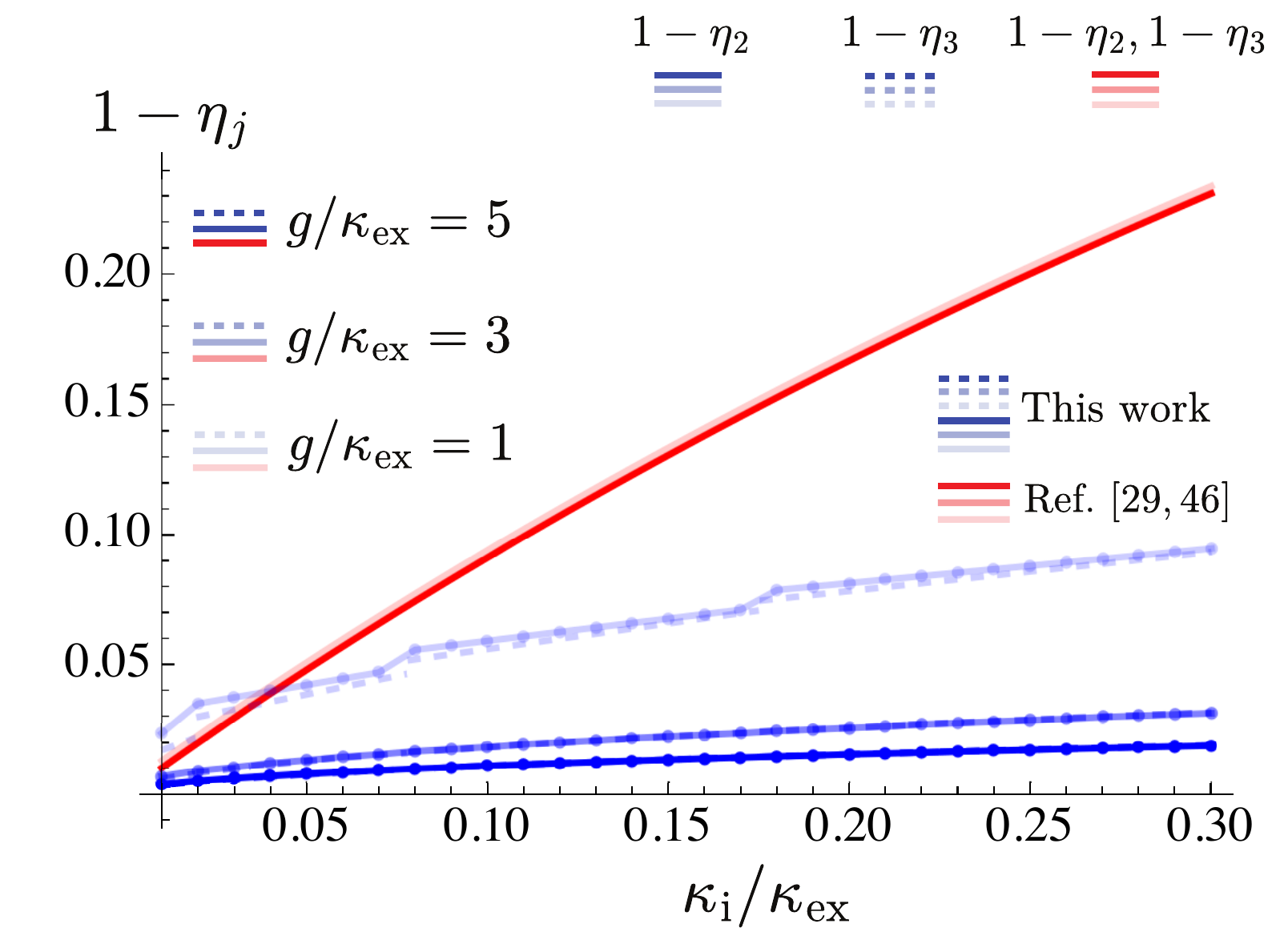}
	\caption{The inefficiency $1-\eta_2$ and single-photon loss $1-\eta_3$ for our protocol are plotted for various values of $\ki/\kx$, with $\kx$ fixed and $\gamma = 0.003 \kx$. Here, the $\mc{T}$ operations are omitted, $\eta_2 = \max_\tau \lvert\int_{-\infty}^\infty a_{\text{out}}(t) h^*(t+\tau) dt \rvert^2$, and $\eta_3 = \int_{-\infty}^\infty \abs{a_{\text{out}}(t)}^2dt $. The dashed-blue lines show the minimum single-photon loss in our protocol from Eq.~(\ref{plossEq}), while the solid-red lines, which overlap, correspond to the maximum achievable values of both $\eta_2$ and $\eta_3$ obtained using the single-strong-pulse protocol \cite{made, vqed} in the limit $\kx T \gg 1$. The inefficiency for our protocol, shown by the solid-blue lines, is computed by numerically solving Eqs.~(\ref{dyn1})--(\ref{dyn3}) with the single-photon mode $h(t) =\sqrt{8/(3T)}\sin^2(\pi t/T)$ and $\kx T = 100$. We use the value of $k$ that minimizes the single-photon loss in Eq.~(\ref{plossEq}).} 
	\label{lossplot}
\end{figure}

We can compare the single-photon loss of our protocol with that in the single-strong-pulse protocol, which has no bound on the control-pulse size. 
For the single-strong-pulse protocol, in the limit $\kx T \gg 1$, both the inefficiency $1-\eta_2$ and the single-photon loss $1-\eta_3$ attain the same minimum value, $1-\eta_{\text{max}}$, where $\eta_{\text{max}}$ is given in Eq.~(\ref{effc}) \cite{made, vqed}.
In that case, the single-photon loss is approximately $1 - \eta_3 \approx (\kx C^{-1})/\kappa + \ki/\kappa$ for $C \gg 1$. In the limit $C_{\text{ex}}^{-1} \ll (1, \ki/\kappa)$, the single-photon loss for the single-strong-pulse protocol is approximately $\ki/\kappa$, since $C^{-1} \leq 2 C_{\text{ex}}^{-1}$ [from $\ki <\kx$ in the discussion following Eq.~(\ref{dyn3})]. We now consider our protocol for which $1-\eta_3$ can be approximated, for $k \gg 1$, as $ \pi^2 \ki / (16k\kx) + 4kC_{\text{ex}}^{-1}$. In this case, the single-photon loss depends on the parameters $(C_{\text{ex}},\ki/\kx)$ and has a minimum at $k = \frac{\pi}{8}\sqrt{\ki C_{\text{ex}}/\kx}$ to yield
\begin{align}
1-\eta_3^{\text{opt}} &\approx \pi \sqrt{\frac{\ki}{\kx C_{\text{ex}}}},
\end{align}
where $\eta_3^{\text{opt}}$ is the maximal value of $\eta_3$. As shown in Fig.~\ref{lossplot}, for our protocol with single-photon mode $h(t) =\sqrt{8/(3T)}\sin^2(\pi t/T)$ and $\kx T = 100$, the single-photon loss and inefficiency are approximately equal in the limits $C_{\text{ex}}^{-1} \ll 1$ and $\kx/(g^2 T) \ll 1$. Moreover, for a large range of $\ki/\kx$, our protocol using the optimal value of $k$ has higher values of efficiency $\eta_2$ compared to those achievable by the single-strong-pulse protocol. The improvement in efficiency increases with $\ki/\kx$.

\begin{figure}[h]
		\includegraphics[
		width=\columnwidth
		]{ 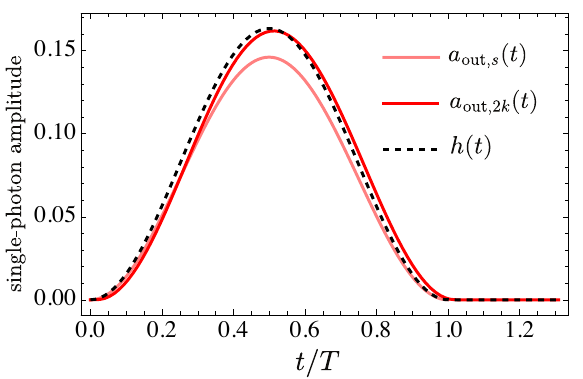}
	\caption{The single-photon output $a_{\text{out}, 2k}(t)$, obtained from numerical solutions of Eqs.~(\ref{dyn1})--(\ref{dyn3}) for $k = 18$, is shown. Here, the $\mc{T}$ operations are omitted, the single-photon mode is $h(t)=\sqrt{8/(3T)}\sin^2(\pi t/T)$, and the system parameters are $(g, \ki, \gamma) = (5, 0.25, 0.003)\kx$, with $T = 100/\kx$. The single-photon output $a_{\text{out}, s}(t)$, obtained from the single-strong-pulse protocol \cite{made, vqed}, is shown for comparison. The inefficiency for our protocol is $1-\eta_2 = 0.017$, where $\eta_2 = \max_\tau \lvert\int_{-\infty}^\infty a_{\text{out}}(t) h^*(t+\tau)\, dt \rvert^2$. The inefficiency for the single-strong-pulse protocol is $1-\eta_2 = 0.2$. The $y$-axis has units of $T^{-1/2}$.} 
	\label{loss12}
\end{figure}

For the example of single-photon retrieval shown in Fig.~\ref{loss12}, with $\ki/\kx = 0.25$, $g=5\kx$, $\gamma = 0.003 \kx$, $h(t) =\sqrt{8/(3T)}\sin^2(\pi t/T)$, and $\kx T = 100$, the single-strong-pulse protocol has inefficiency $1-\eta_2= 0.2$, whereas the amplification protocol without applying $\mathcal{T}$ operations gives $1-\eta_2 = 0.017$ for $k=18$. Our protocol therefore motivates using amplitude amplification for
error reduction
in physical setups with a lossless channel (the waveguide, in our case), allowing the desired amplitude to be built iteratively without substantially populating the lossy channel (the cavity with intrinsic loss). Additionally, if $(1-\epsilon_t)$ is the transmission probability per trip through the waveguide, in the limits $ k \gg 1 $ and $ \epsilon_t k \ll 1$, we can upper bound the single-photon loss through the waveguide by $ \approx \epsilon_t (k + 3/2)$. See Appendix \ref{LossDerivation} for more details.

\subsection{Fock-state preparation in superconducting circuit QED} \label{res1}

\begin{figure}[h]
		\includegraphics[
		width=0.5
        \columnwidth
		]{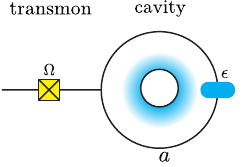}
    \caption{ Schematic diagram of a transmon coupled to a cavity mode $ a$.}
	\label{Transmon}
\end{figure}

In this subsection, we consider a superconducting qubit coupled to a single mode of a high-quality-factor cavity, as shown in Fig.~\ref{Transmon}, for Fock-state generation. The system in the dispersive limit is described by the following Hamiltonian \cite{Schuster_2007, PhysRevA.92.040303}:
\begin{align}
H_0=\omega_q \ketbra{e}{e}+\omega_c   \hat n - \chi \ketbra{e}{e}  \hat n, \label{h0cir}
\end{align}
where $\ket{e}, \ket{g}$ are the qubit states, $\omega_q$ is the qubit transition frequency between $\ket{g}$ and $\ket{e}$, $\omega_c$ is the cavity frequency, $ a^\dagger$ ($ a$) are the creation (annihilation) operators for the cavity mode, $ \hat n =  a^\dagger  a$ is the photon number operator, and $\chi$ is the dispersive coupling between the transmon and the cavity. The above Hamiltonian can be derived from the transmon-cavity Hamiltonian in the limit where $g \ll \abs{\Delta}$ and $g \ll \abs{\Delta+\tilde \alpha}$, where $\Delta  = \omega_q -\omega_c$, $g$ is the coupling between the first two levels of the transmon and the cavity, and $\tilde \alpha$ represents the anharmonicity of the transmon \cite{PhysRevA.76.042319}.

As shown in Ref.~\cite{PhysRevA.92.040303}, implementing the Hamiltonians $H_0 + H_1$ and $H_0 + H_2$, where
\begin{align}
H_1 &= \mc{E} e^{-i\omega_c t} a^\dagger + \text{H.c.},\label{h1cir}\\
H_2 &= \Omega(t)e^{-i\omega_qt}\ketbra{e}{g} + \text{H.c.}\label{h2cir},
\end{align}
is sufficient for universal control of the oscillator. Here, $H_2$ is written using the rotating-wave approximation. When the photon mode is driven with the qubit in the state $\ket{g}$, the Hamiltonian  $H_0 +H_1$ implements the displacement
\begin{align}
 D(\lambda) = \exp(\lambda  a^\dagger - \lambda^* a),
\end{align}
where $\lambda = -i \mc{E} \bar{T}$, with $\bar{T}$ the evolution time.
Moreover, the Hamiltonian $H_0 + H_2$ implements the selective number-dependent arbitrary phase (SNAP) gate \cite{PhysRevLett.115.137002}, which is composed of a sequence of unitaries
$ S_n(\theta_n)= e^{i\theta_n \ketbra{n}{n}}$. The SNAP gate $\prod_i S_i(\theta_i)$ can be implemented by driving the qubit, initialized in the state $\ket{g}$, with $\Omega(t)=\sum_{n=0}^{\infty}\Omega_n(t)e^{i n \chi t}$, where $\abs{\Omega_n(t)} \ll \chi$ for all $n$. Here, $\theta_n$ depends on the closed path traversed on the Bloch sphere corresponding to $\{\ket{g,n}, \ket{e,n} \}$. Moreover, each $S_n(\theta_n)$ can be implemented using the sequence of rotations $U_{\text{SNAP}} = R( \phi_{ 1})R( \phi_{2}) = -\exp[i(\phi_{ 2} - \phi_{1})\sigma_z]$, where $R(\phi) = \exp[-i\pi/2(\cos \phi \sigma_x + \sin\phi \sigma_y)]$, giving $\theta_n = \phi_{ 2}- \phi_{ 1}+\pi$. Here, the Pauli matrices are written in the basis $\{\ket{g, n}, \ket{e, n} \}$, and the Bloch-sphere rotation $R(\phi )$ is implemented by time evolution under $H_0+H_2$, with $\Omega(t) = \abs{\Omega_n(t)}e^{i\phi}e^{i n\chi t}$, $\int_0^{\bar T} dt|\Omega_n(t)| = \pi/2$, and $\bar T$ the evolution time.

\begin{figure}[h]
		\includegraphics[
		width=\columnwidth
		]{ 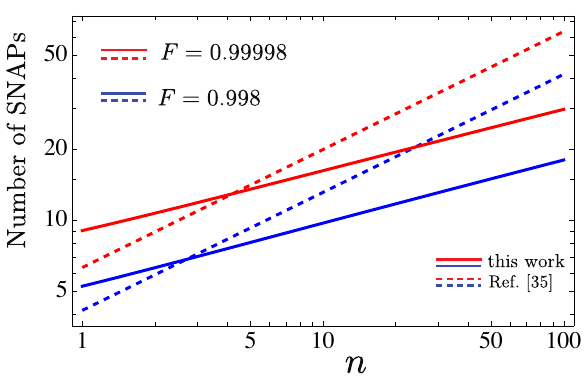}
	\caption{We show the number of SNAP gates, given by $2l$ from Eq.~(\ref{lbound}), required to obtain the Fock state $\ket{n}$ with different fidelities. The axes are plotted using logarithmic scaling. For a given state fidelity, the solid line corresponds to our protocol, while the dashed line shows a numerical fit to the gate count of the protocol in Ref.~\cite{PhysRevA.92.040303}. The red lines correspond to the state fidelity $F = 0.99998$, while the blue lines correspond to the state fidelity $F = 0.998$.} 
	\label{SNAPC}
\end{figure}

We now outline our Fock-state preparation protocol based on fixed-point amplitude amplification and describe the implementation of $U_{G_1}$ from Eq.~(\ref{G2fp}). We take the initial cavity state to be the coherent state $\ket{\psi_0} =  D(\lambda)\ket{0}=\sum_{m=0}^{\infty}c_m(\lambda) \ket{m}$, where $\lambda = \sqrt{n}$, with the transmon initialized in $\ket{g}$. We first construct the modified reflection operator about $\ket{n}$, which we want to prepare, i.e., $ R_1(\beta) = I - (1 - e^{i \beta}) \ketbra{n}{n}$. This unitary is implemented using the SNAP gate $
S_n(\beta) = e^{i \beta \ketbra{n}{n}}$, so that $R_1(\beta) = S_n(\beta)$. We then construct the modified reflection about the initial state $R_i(\alpha)$ as follows:
\begin{align}
R_i(\alpha)&= I - (1-e^{i\alpha})\ketbra{\psi_0}{\psi_0} \label{rideffock}\\
&= D(\lambda) [I - (1-e^{i\alpha})\ketbra{0}{0}] D(\lambda)^\dagger\\
&=D(\lambda) S_0(\alpha) D(\lambda)^\dagger, \label{ria}
\end{align}
where $S_0(\alpha) = e^{i\alpha \ketbra{0}{0}}$, i.e., the modified reflection about the initial state $D(\lambda)\ket{0}$ can be implemented using two cavity displacements and a SNAP gate. The unitary sequence that implements fixed-point amplification for preparing the Fock state $\ket{n}$ is given by
\begin{align}
U_n = (-1)^l \prod_{j=1}^{l}D(\lambda) S_0(\alpha_j) D(\lambda)^\dagger S_n(\beta_j) \label{Un},
\end{align}
where the unitaries are ordered from right to left.
Finally, using Eq.~(\ref{lbound}) and the fact that $\bra{n}D(\lambda)\ket{0} =  e^{-n/2}n^{n/2}/\sqrt{n!} \approx 1/(2\pi n)^{1/4}$ for $n \gg 1$, we can choose phases in the $S_i$ in $U_{G_1}$ to produce the Fock state $\ket{n}$ using $\lceil 2l \rceil$ SNAP and displacement gates, where 
\begin{align}
2l
& = (2\pi)^{1/4}\log(2/\delta)\bigr(n^{1/4} + O(1/n)\bigr) -1, \label{2ldrel}
\end{align}
and $1-\delta^2$ is the success probability from Eq.~(\ref{lbound}). This gives a quadratic improvement from $O(n^{1/2})$ SNAP gates in Ref.~\cite{PhysRevA.92.040303}. Each amplification step in $U_n$ requires two SNAP gates and two displacement gates. In Fig.~\ref{SNAPC}, we show the required number of SNAP gates versus $n$ for different values of the state fidelity $F=\abs{\braket{n}{\psi_{\text{amp}}}}^2$, where $\ket{\psi_{\text{amp}}}$ is the state obtained from applying $U_n$ to the coherent state $D(\sqrt{n})\ket{0}$. 
We observe that, for practical values of the fidelity and photon number \cite{Deng2024}, our protocol reduces the required number of SNAP gates compared to Ref.~\cite{PhysRevA.92.040303}.

\subsubsection*{NOON-state preparation}
\begin{figure}[h]
		\includegraphics[
		width=0.52
        \columnwidth
		]{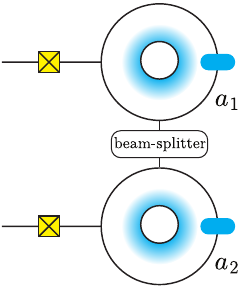}
    \caption{We present a schematic diagram of two transmon–cavity systems with modes $a_1$ and $ a_2$ used for NOON state preparation. A beam-splitter interaction is used to implement $U_{\text{BS}}$ on the cavity modes from Eq.~(\ref{bsuse}).}
	\label{transmon3}
\end{figure}

A variation of the approach used in
Eq.~(\ref{Un})
can be implemented in multimode circuit QED with two cavity modes \cite{MultiMode} to prepare the NOON state \cite{PhysRevA.65.052104} $\ket{\psi_{\text{NOON}}} \propto \ket{0,n}_{1,2} + \ket{n,0}_{1,2}$, where the subscripts label the cavity modes. As shown in Fig.~\ref{transmon3}, we have two transmon-cavity systems as described in Eqs.~(\ref{h0cir})--(\ref{h2cir}). We use the initial photonic state $\ket{\psi_0} \propto \ket{0, \lambda}_{1,2} + \ket{\lambda, 0}_{1,2}$ with $\lambda = \sqrt{n}$.
We can prepare $\ket{\psi_0}$ as follows: we prepare the cat state \cite{RevModPhys.85.1103, catstate1, catstate2} in the first mode and the vacuum state in the second mode to obtain $(|-\sqrt{2}\lambda/2 \rangle_1+|\sqrt{2}\lambda/2 \rangle_1)\ket{0}_2$. We then apply the cavity displacement $D(\sqrt{2}\lambda/2)$ on the first mode to obtain the state $(\ket{0}_1+|\sqrt{2}\lambda \rangle_1)\ket{0}_2$.
We then apply the beam-splitter unitary $U_{\text{BS}}$ \cite{bs0, bs1, bs2, PRXQuantum.4.020355} such that $U_{\text{BS}}a_1 U_{\text{BS}}^\dagger = (a_1-a_2)/\sqrt{2}$ and $U_{\text{BS}}a_2U_{\text{BS}}^\dagger = (a_1+a_2)/\sqrt{2}$. We can then use the fact that
\begin{align}
U_{\text{BS}}D_1(\sqrt{2}\lambda)\ket{0, 0}_{1,2}&=D_1(\lambda)D_2(-\lambda)U_{\text{BS}}\ket{0, 0}_{1,2}, \label{bsuse}
\end{align}
where $D_i$ denotes the displacement on the $i$th mode, and $U_{\text{BS}}\ket{0, 0}_{1,2}=\ket{0,0}_{1,2}$, to obtain the state $\ket{0, 0}_{1,2}+\ket{\lambda, -\lambda}_{1,2}$. Finally, applying the displacement $D_2(\lambda)$ gives us $\ket{\psi_0}$.

\begin{figure}[t]
		\includegraphics[
		width=\columnwidth
		]{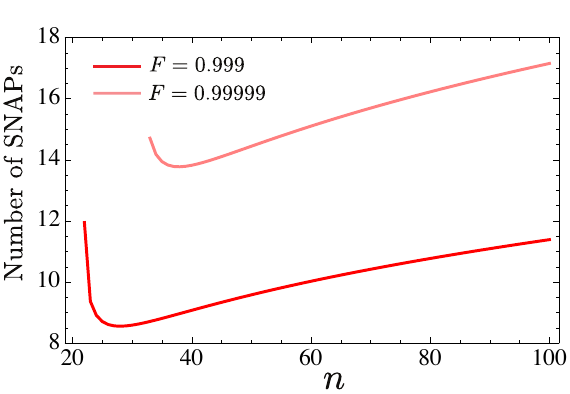}
	\caption{We show the number of SNAP gates, $2l$, required for NOON state preparation, obtained by numerically solving Eq.~(\ref{fnoon}) for different values of the state fidelity $F$.} 
	\label{FNOON}
\end{figure}

As in Eq.~(\ref{Un}), we now denote $U_n^{(i)}$ as the unitary sequence $U_n$ to prepare the Fock state $\ket{n}_i$  in the $i$th mode. Here, the modified reflections $R_i(\alpha)$ and $R_1(\beta)$ in $U_n^{(i)}$ are implemented using the SNAP and displacement gates on the $i$th mode using the dispersive interaction of the cavities with their respective qubits (as previously discussed for Eq.~(\ref{Un})). As shown in Appendix~\ref{errorNOON}, the fidelity $F_{\text{NOON}}$ of the state $U_n^{(2)}U_n^{(1)} \ket{\psi_0}$ satisfies the following lower bound:
\begin{align}
F_{\text{NOON}} \geq 1-2[y +e^{-n}/2],\label{fnoon}
\end{align}
where $F_{\text{NOON}}=\lvert \bra{\psi_{\text{NOON}}}U_n^{(2)}U_n^{(1)} \ket{\psi_0} \rvert^2$, and $y = \delta^2 + \epsilon +\sqrt{2\epsilon}\delta - \epsilon \delta^2$. Here, $\delta$ is the error in the amplification protocol $U_n^{(i)}$ from Eq.~(\ref{2ldrel}), and $\epsilon \leq 2l e^{-n/2}$, where $l$ is the length of the protocol $U_n^{(i)}$. The error has a complicated form because, in  $U_n^{(j)}U_n^{(i)}\ket{0, \lambda}_{i,j}$, we have $U_{n}^{(i)}\ket{0}_i \neq \ket{0}_i$. Choosing a small value of $1-\lvert \bra{n}_iU_n^{(i)}\ket{\lambda}_i \rvert^2 $ requires a large $l$, while the error $\lVert U_n^{(i)}\ket{0}_{i} - \ket{0}_{i} \rVert_2$ increases with $l$. Since $\delta$ depends on $l$ through Eq.~(\ref{lbound}), we can write the error in terms of variables $(l, n)$ and find the optimum scaling of $l$ with $n$. As shown in Appendix \ref{errorNOON}, for $n\gg 1$, the optimum $l$ such that the error falls at the fastest rate is given by $l_{\text{opt}} = cn^{5/4}/4$, where $c = (2\pi)^{1/4}/2$. The error with the optimal $l(n) = l_{\text{opt}}$ gives $1-F_{\text{NOON}} \lesssim  e^{-n/2}n^{5/8}(4\sqrt{c}+n^{5/8}c)$. As shown in Fig.~\ref{FNOON}, we can also use Eq.~(\ref{fnoon}) to numerically compute $l(n)$ to achieve a given lower bound on the fidelity. Moreover, as shown in Fig.~\ref{FNOON}, the maximum achievable fidelity increases with the photon number $n$.
The advantage of our NOON state preparation protocol is that it does not require resonant transmon-cavity interactions \cite{PhysRevLett.115.137002, FastSimpleNoon} or a three-level transmon structure \cite{FastSimpleNoon, PhysRevA.107.042412, Merkel_2010}.
 
All protocols in this subsection rely on SNAP and displacement gates and therefore inherit their implementation errors. As discussed in Ref.~\cite{PhysRevA.92.040303}, the finite value of $|\Omega_n/\chi|$ introduces deviation from the ideal SNAP gate. However, recent works based on optimized pulses $\Omega_n$ \cite{landgraf2024fastquantumcontrolcavities} and unitary sequences derived from quantum signal processing \cite{fong2025engineering} have shown how to implement SNAP gates faster while reducing gate infidelity.

\section{Conclusion and outlook}\label{conout}
In this work, we describe state-preparation protocols based on amplitude amplification in cavity QED for two physical setups, enabling the generation of useful quantum states of light, including traveling single photons, Fock states, and NOON states. Our proposed method, \textit{very oblivious amplitude amplification}, simplifies the unitary used in oblivious amplitude amplification under mild assumptions, yielding a simpler state-preparation protocol for traveling single photons. Since the construction is general, it can also be applied to other quantum algorithms and state-preparation tasks that rely on amplitude amplification. We anticipate our work will be a stepping stone to practical and scalable protocols for preparing other useful and important quantum states for applications in quantum computation, quantum communication, and quantum metrology, including code states for bosonic quantum error-correcting codes \cite{PhysRevA.64.012310,Mirrahimi_2014}. For state preparation in circuit QED systems, we restrict attention to unitary sequences of SNAP and displacement gates, for which we provide analytic gate decompositions and fidelity guarantees. Extending these results to other Hamiltonian interactions \cite{WeakFock} with different gate sets remains an interesting open direction.
Additionally, we leave the task of preparing traveling photonic Fock states with higher photon numbers as an open problem. We anticipate that using amplitude amplification in this setting will require developing the analog of SNAP gates for traveling photons.\\[1em]
\begin{acknowledgements}
S.A.~thanks Mauro E.~S.~Morales, Sagar Airen, Sagnik Mondal, and Rajrupa Mondal for helpful discussions.
S.A., Z.Y.W., and A.V.G.~were supported in part by ARL (W911NF-24-2-0107), DARPA SAVaNT ADVENT, NSF QLCI (award No.~OMA-2120757), DoE ASCR Quantum Testbed Pathfinder program (award No.~DE-SC0024220), ONR MURI, NSF STAQ program, AFOSR MURI, and NQVL:QSTD:Pilot:FTL. S.A., Z.Y.W., and A.V.G.~also acknowledge support from the U.S.~Department of Energy, Office of Science, National Quantum Information Science Research Centers, Quantum Systems Accelerator (award No.~DE-SCL0000121) and from the U.S.~Department of Energy, Office of Science, Accelerated Research in Quantum Computing, Fundamental Algorithmic Research toward Quantum Utility (FAR-Qu). K.S. acknowledges support from Northrop Grumman (seedling grant) and the DARPA SAVaNT program (through ARO award No. W911NF2120106).
\end{acknowledgements}

\bibliographystyle{apsrev4-2}
\bibliography{Ref2}

\appendix

\section{Implementing $\bar U^\dagger$ in the emitter--cavity system} \label{ufder}

In this section, we derive the implementation of the unitary $\bar U^\dagger$, where $\bar U$ is given in Eq.~(\ref{apstate}). Moreover, we ignore all losses in the system, i.e., $\gamma = 0$ and $\kappa = \kx$. Let $\Omega(t)$ implement the dynamics in Eqs.~(\ref{dyn1})--(\ref{dyn3}) and effect the unitary $\bar U$ on the initial state $\ket{s,0_p}$:
\begin{align}
\bar U\ket{s, 0_p} = \cos \theta \ket{s, 0_p}+\sin\theta \ket{g, 1_h},
\end{align}
where $\ket{1_h}$ denotes the outgoing single photon with mode $h(t)$. Expressions for $\Omega(t)$ can be found in \cite{made,gauss,vqed}. The equations that govern the dynamics are given as follows:
\begin{align}
\dot{c}_s&= -i\Omega(t)^* c_e, \label{d1}\\
\dot{c}_e &= -i\Omega(t)c_s -ig c_g, \label{d2}\\
\dot{c}_g &= -ig c_e - \kappa c_g -\sqrt{2\kappa}a_{\text{in}},\label{d3}
\end{align}
with $a_{\text{out}} = a_{\text{in}} + \sqrt{2\kx}c_g$. Here $(c_s, c_e, c_g )$ denote amplitudes in the states $(\ket{s, 0}, \ket{e, 0}, \ket{g, 1})$.

We now show that the unitary $U_b$ effected by the control pulse $\Omega_R(t) = -\Omega(T-t)$ in Eqs.~(\ref{dyn1})--(\ref{dyn3}) with the input field $a_{\text{in}}^{R}(t) =\mc{T}a_{\text{out}}(t) \mc{T}= a_{\text{out}}(T-t)$ implements the following:
\begin{align}
\mathcal{T}U_b \mathcal{T}  \bar U \ket{\phi} = \ket{\phi}, \label{Udeqmain}
\end{align}
where $\ket{\phi}$ is any state in the subspace $\{\ket{s, 0}, \ket{g, 1_{\bar f}} \}$, $a_{\text{in}}(t)$ is the single-photon input in $\ket{\phi}$ with an arbitrary mode function $\bar f$, and $\mathcal{T}$ represents the time reversal operation on the optical mode $f(t)$ given as $f(t)\rightarrow f(T-t)$.
With $\ket{\phi}$ as the initial state, the dynamics implemented by $\bar U$ are represented by state amplitudes $(c_s(t), c_e(t), c_g(t))$ that obey Eqs.~(\ref{d1}--\ref{d3}), and the output single-photon mode is $a_{\text{out}}(t) = a_{\text{in}}(t)+\sqrt{2\kx}c_g(t)$. As we will now show, $\bar U^\dagger$  can be implemented using $\mathcal{T}U_b \mathcal{T}$. We first show that the evolution $ U_b  $ has state amplitudes $(c_s^R(t), c_e^R(t), c_g^R(t)) = (c_s(T-t), c_e(T-t), -c_g(T-t))$, with $a_{\text{in}}^R(t) = a_{\text{out}}(T-t)$ as the single-photon input.
For $\dot{c}_s^{R}$, we have the differential equation
\begin{align}
\dot c_s^R &= -\dot c_s(T-t)\\
&=i\Omega(T-t)^* c_e(T-t)\\
&=-i\Omega_R(t)^* c_e^R(t),
\end{align}
where we use Eq.~(\ref{d1}), $c_s(T-t) = c_s^R(t)$, and $\Omega_R(t)=-\Omega(T-t)$. Similarly, for $c_e^R(t)$, we have the differential equation
\begin{align}
\dot c_e^R &= -\dot c_e(T-t)\\
&=-\bigb{-i\Omega(T-t)c_s(T-t)-igc_g(T-t)}\\
&=-i\Omega_R(t)c_s^R(t)-igc_g^R(t),
\end{align}
where we use Eq.~(\ref{d2}) and $c_e^R(t) = c_e(T-t)$.
Finally, for $c_g^R(t)$, we have the differential equation
\begin{align}
\dot c_g^R=& \dot c_g(T-t)\\
=&-igc_e(T-t) - \kappa c_g(T-t) - \sqrt{2\kappa}a_{\text{in}}(T-t)\label{t1}\\
=&-igc_e^R(t)+\kappa c_g^R(t) - \sqrt{2\kappa}\bigb{a_{\text{out}}(T-t) \label{t2}\\&- \sqrt{2\kappa}c_g(T-t)} \notag\\
=& -ig c_e^R(t) +\kappa c_g^R(t) - \sqrt{2\kappa}\bigb{a_{\text{out}}(T-t)\\&+\sqrt{2\kappa}c_g^R(t)}\\
=&-ig c_e^R(t) - \kappa c_g^R(t) - \sqrt{2\kappa}a_{\text{out}}(T-t)\\
=& -ig c_e^R(t) - \kappa c_g^R(t) - \sqrt{2\kappa}a_{\text{in}}^R(t) \label{last}.
\end{align}
In Eq.~(\ref{t1}), we used Eq.~(\ref{d3}). In Eq.~(\ref{t2}), we used the facts that $c_g^R(t) = -c_g(T-t)$ and  $a_{\text{out}}(t) = a_{\text{in}}(t) + \sqrt{2\kappa}c_g(t)$. In Eq.~(\ref{last}), we used $a_{\text{in}}^{R}(t) = a_{\text{out}}(T-t)$. To summarize, the state amplitudes $(c_s^R, c_e^R, c_g^R)$ satisfy the following equations:
\begin{align}
\dot c_s^R &= -i\Omega_R(t)^* c_e^R(t),\\
\dot c_e^R &= -i\Omega_R(t)c_s^R(t)-igc_g^R(t),\\
\dot c_g^R &= -ig c_e^R(t) - \kappa c_g^R(t) - \sqrt{2\kappa}a_{\text{in}}^R(t),
\end{align}
which correspond to dynamics in Eqs.~(\ref{dyn1})--(\ref{dyn3}) with control pulse $\Omega_R(t)$ and single-photon input $a_{\text{in}}^{R}(t)$. Moreover, the single-photon output $a_{\text{out}}^{R}(t)$ is
\begin{align}
a_{\text{out}}^{R}(t) &= a_{\text{in}}^R(t) + \sqrt{2\kx}c_g^R(t)\\
&= a_{\text{out}}(T-t) - \sqrt{2\kx}c_g(T-t)\\
&= a_{\text{out}}(T-t)- [a_{\text{out}}(T-t) - a_{\text{in}}(T-t)]\\
&= a_{\text{in}}(T-t).
\end{align}
We first note that $\bar U \ket{\phi}$ is characterized by the amplitudes $c_s(T)$ and $a_{\text{out}}(t)$, with all other amplitudes vanishing. Moreover, $U_b$ maps the inputs $c_s^R(0) = c_s(T)$ and $a_{\text{in}}^R(t) = a_{\text{out}}(T-t)$ to the outputs $c_s^R(T) = c_s(0)$ and $a_{\text{out}}^R(t) = a_{\text{in}}(T-t)$. Therefore, $\mc T U_b \mc T$ on the state $\bar U \ket{\phi}$ yields the outputs $c_s(0)$ and $a_{\text{in}}(t)$, i.e., the state $\ket{\phi}$, and hence implements $\bar U^\dagger$.

Moreover, in the Grover unitary $G_3 = \bar U \bar R_1 \bar U^\dagger \bar R_1 = \bar U \bar R_1 \mathcal{T}  U_b\mathcal T \bar R_1  $, we can absorb $\bar R_1$ into $U_b$ resulting in a $\pi$ phase shift on the control pulse $\Omega_R(t)$ that implements $U_b$. This then gives us the Grover unitary $G_3 = \bar U \mathcal T W \mathcal T$, where $W$ is effected using the control pulse $\Omega_W(t)=\Omega(T-t)$. Moreover, for an analogous calculation with a different system where the input-output relations are given as \cite{alexeyP1} 
\begin{align}
\dot{a}&=-i[a, H] -\kx a + \sqrt{2\kx} a_{\text{in}}(t),\\
a_{\text{out}}(t) &= -a_{\text{in}}(t) + \sqrt{2\kx}a(t),
\end{align}
which correspond to a single-sided cavity geometry, we use $\Omega_W(t) = -\Omega(T-t)$ instead.
\section{Analytic solution for $\Lambda$-type cavity QED single-photon generation for long Grover sequences}\label{appbee}

In this section, we consider the case $k \gg 1$ with $\theta = \pi/[2(2k+1)]$ in Eq.~(\ref{apstate}). Here, $k$ denotes the length of the Grover sequence from Eq.~(\ref{ug3}), and we derive analytic expressions for the state evolution generated by $U_{G_3}$ from Sec.~\ref{cqed}.

We consider the following system of equations governing the dynamics of the physical system for an arbitrary control pulse $\Omega (t)$ with an input pulse $a_{\text{in}}(t)$ and $a_{\text{out}}(t) = a_{\text{in}}(t) + \sqrt{2\kx}c_g(t)$:
\begin{align}
\dot{c}_s&= -i\Omega(t)^* c_e, \label{e1}\\
\dot{c}_e &= -i\Omega(t)c_s -ig c_g - \gamma c_e, \label{e2}\\
\dot{c}_g &= -ig c_e - \kappa c_g -\sqrt{2\kx}a_{\text{in}},\label{e3}
\end{align}
where $\kappa = \kx+\ki$.
We now consider the case where the control pulse $\Omega(t)$ is weak and can be written as $\Omega(t) = \epsilon \zeta(t)$, where $\epsilon \ll 1$. We now expand the solutions in powers of $\epsilon$ as follows: 
\begin{align}
c_s = c_s^{(0)} + \epsilon c_s^{(1)} + \epsilon^2 c_s^{(2)} + \cdots,\\
c_e = c_e^{(0)} + \epsilon c_e^{(1)} + \epsilon^2 c_e^{(2)} + \cdots,\\
c_g = c_g^{(0)} + \epsilon c_g^{(1)} + \epsilon^2 c_g^{(2)} + \cdots.
\end{align}
We first write the zeroth-order differential equations as follows:
\begin{align}
\dot{c}_s^{(0)} & = 0, \label{ord01}\\
\dot{c}_e^{(0)} & = -ig c_g^{(0)}- \gamma c_e^{(0)},\label{ord02}\\
\dot{c}_g^{(0)} & = -ig c_e^{(0)} - \kappa c_g^{(0)} - \sqrt{2\kx}a_{\text{in}}(t).\label{ord03}
\end{align}
These equations can be solved in the adiabatic limit [$\kx/(g^2 T) \rightarrow 0$] as follows:
\begin{align}
c_s^{(0)}(t) &= c_s(0) ,\\
c_e^{(0)}(t) &= \Big( \frac{C}{1+C}\Big)\frac{i\sqrt{2\kx}}{g}a_{\text{in}}(t),\\
c_g^{(0)}(t) &= \frac{i\gamma}{g}c_e^{(0)}. \label{cg0sol}
\end{align}
The differential equations for order $j\geq 1$ are as follows:
\begin{align}
\dot{c}_s^{(j)} & = -i\zeta^*(t)c_e^{(j-1)}, \label{cs1diff}\\
\dot{c}_e^{(j)} & = -i\zeta(t) c_s^{(j-1)}- \gamma c_e^{(j)}  - igc_g^{(j)},\\
\dot{c}_g^{(j)} & = -ig c_e^{(j)} - \kappa c_g^{(j)}(t).\label{cg1diff}
\end{align}
The above equations can be solved in the adiabatic limit [$\kx/(g^2 T) \rightarrow 0$] to give the relation
\begin{align}
c_g^{(j)}(t) &= - \Big(\frac{C}{ 1+C }\Big)\frac{\zeta (t)}{g}c_s^{(j-1)}(t), \>\>\> j \geq 1\label{cgjrel},
\end{align}
and for $j=1$,
\begin{align}
c_s^{(1)}(t) & = \Big(\frac{C}{ 1+C }\Big) \frac{\sqrt{ 2\kx}}{g} \int_0^t \zeta^*(\tau)a_{\text{in}}(\tau)d\tau \label{cs1Sol},
\end{align}
where $C = g^2/(\kappa \gamma)$, and we use Eq.~(\ref{cs1diff}) and Eq.~(\ref{cg0sol}). We can then use Eq.~(\ref{cs1Sol}) and Eq.~(\ref{cgjrel}) to obtain the solution
\begin{align}
a_{\text{out}}(t)=& a_{\text{in}}(t) + \sqrt{2\kx}\Big(c_g^{(0)}(t) + \epsilon c_g^{(1)}(t) \notag \\&
+ \epsilon^2 c_g^{(2)}(t)   \Big)\\
=& a_{\text{in}}(t) - \frac{2\kx}{\kappa(1+C)}a_{\text{in}}(t) \notag \\ &- \frac{\sqrt{2\kx}\zeta(t)}{g(1+C^{-1})}\Big(c_s(0)\epsilon + c_s^{(1)}(t)\epsilon^2  \Big).  \label{lastaout}
\end{align}

We now use the above results for our problem, in which the amplification protocol $U_{G_3} = S_{2k}\hdots S_2S_1S_0$, given in Eq.~(\ref{ug3}), is applied to the initial state $\ket{s, 0_p}$.
We denote $a_{\text{out},j}(t)$ as the single-photon output obtained at the end of step $S_j$. We consider the case where $\gamma = \kappa_{\text{i}}=0$, i.e., the process is unitary. First, note that the control pulse $\Omega(t)= \epsilon \zeta(t)$ that implements the unitary $\bar U$  has \cite{made,vqed}
\begin{align}
\zeta(t) &= -\frac{g}{\sqrt{2\kx}}h(t),
\end{align}
with $\epsilon = \sin\theta$.
From results in Appendix \ref{ufder}, the control pulse that implements the unitary $W$ is $\Omega(t)$ as well since $h(t) = h(T-t)$. We can then write a recursion relation between the single-photon output $a_{\text{out},j}(t)$ and the amplitude of state $s$ at the end of step $j$,  $c_{sj}(T)$, as follows:
\begin{align}
a_{\text{out}, j}(t) &= a_{\text{out}, j-1}(t) + \sin\theta h(t)c_{sj-1}(T), \label{recur1}\\
c_{sj}(T)&= c_{sj-1}(T)-\sin\theta \int_0^T h(t^\prime)a_{\text{out}, j-1}(t^\prime) dt^\prime,\label{recur2}
\end{align}
with the initial conditions $a_{\text{out},0}(t)= h(t)\sin\theta$ and $c_{s0}(T)=\cos\theta$, and $\theta$ is related to $k$ using Eq.~(\ref{thetak}). The solution to this can be found by observing that $a_{\text{out},j}(t)=x_jh(t)$ to obtain the following recursion relation:
\begin{align}
x_j &= x_{j-1}+\sin\theta y_{j-1},\\
y_j &=y_{j-1}-\sin\theta x_{j-1},
\end{align}
where $y_j:=c_{sj}(T)$. The two equations can be combined to form the relation $z_j = (1-i\sin\theta)z_{j-1}$, where $z_j = x_j +iy_j$ and $z_0 = ie^{-i\theta}$, giving us
\begin{align}
a_{\text{out},j}(t)&= r^{j/2}\sin(\theta + j\phi)h(t), \label{ideala}\\
\sqrt{2\kx}c_{gj}(t)&= r^{j/2}\sin\theta \cos[\theta + (j-1)\phi]h(t),
\end{align}
where $r = (1+\sin^2\theta)$, $\phi=\arctan(\sin\theta) \approx \theta$. This shows that the single-photon output at every step of the protocol obeys $a_{\text{out}, j}(t) = a_{\text{out}, j}(T-t)$ which means the $\mathcal{T}$ operations can be omitted.
Moreover, it is easy to check from Eq.~(\ref{ideala}) that the final single-photon output converges to the ideal mode function $h(t)$ for $k\gg1$.

\begin{figure}[t]
		\includegraphics[
		width=\columnwidth
		]{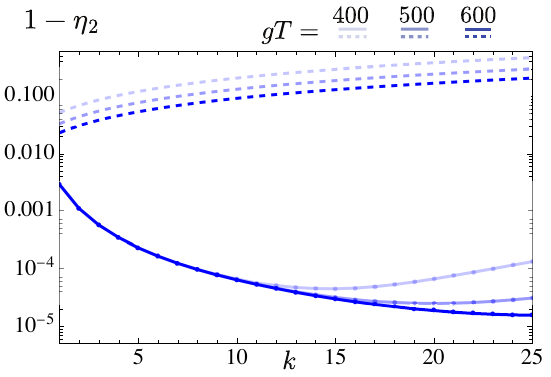} 
    \caption{The inefficiency $1-\eta_2$, where $\eta_2 = \max_\tau \lvert{\int_{-\infty}^\infty a_{\text{out}}(t) h^*(t+\tau) dt }\rvert^2$, is plotted as solid lines against $k$, the length of our protocol in Eq.~(\ref{ug3}), for various values of $gT$ in the lossless case ($\ki = \gamma =0$). The $\mc{T}$ operations are omitted. The inefficiency of our protocol is computed by numerically solving Eqs.~(\ref{dyn1})--(\ref{dyn3}) with the single-photon mode $h(t) =\sqrt{8/(3T)}\sin^2(\pi t/T)$. Here, $T$ is fixed, and $\kx T = 150$. The dashed curves show the corresponding inefficiencies obtained from the upper bound in Eq.~(\ref{worst}). The $y$-axis uses logarithmic scaling.} 
	\label{comparison}
\end{figure}

We note, however, that for finite $\kx T$, the solutions for ($c_{s}^{(j)}$, $c_{g}^{(j)}$) have non-adiabatic corrections of order $\kx/ (g^2T)$ at each step of the protocol. This can be seen from Fourier transforming the equations for the amplitudes and keeping first order terms in $-i\omega$, where $\omega$ is the frequency parameter in the Fourier transform. As shown by numerics, these non-adiabatic corrections lead to deviations from the ideal output mode for increasing $k$. However, the errors can be made infinitesimally small by decreasing $\kx/ (g^2T)$ when increasing $k$. We now analyze the non-adiabatic contribution to the infidelity of our protocol by solving Eqs.~(\ref{e1})--(\ref{e3}) to leading order in $ \kx/(g^2 T)$ for the lossless case $\ki = \gamma = 0$. The zeroth-order amplitudes in Eqs.~(\ref{ord01})--(\ref{ord03}) have the solutions
\begin{align}
c_s^{(0)}(t) &=  c_s(0) ,\\
c_g^{(0)}(t) &= -\frac{\sqrt{2\kx}}{g^2}\dot{a}_{\text{in}},\\
c_e^{(0)}(t) &= \frac{i\sqrt{2\kx}}{g}a_{\text{in}} - \frac{i\kx \sqrt{2\kx}}{g^3} \dot{a}_{\text{in}}.
\end{align}
We can then extend our recursion relations in Eqs.~(\ref{recur1}) and (\ref{recur2}) for single-photon output and atomic state amplitudes, $\bar a_{\text{out}, j}(t)$ and $\bar c_{sj}(T)$, respectively, to obtain
\begin{align}
\bar a_{\text{out}, j}(t) =&\>  \bar a_{\text{out}, j-1}(t)+\sin\theta \bar c_{sj-1}(T)h(t) -\frac{2\kx}{g^2}\dot{\bar a}_{\text{out}, j-1}(t) \notag \\ &- \frac{\kx }{g^2}\bar c_{sj-1}(T)\sin\theta\dot{h}(t)\\
=& \bar a_{\text{out}, j-1}(t-2\beta)+\sin\theta \bar c_{sj-1}(T)h(t-\beta) + O(\beta^2),\\
\bar c_{sj}(T) =& \bar c_{sj-1}(T)\hspace{-0.15em} - \hspace{-0.15em}\sin\theta \int_0^T \hspace{-0.8em} \Big(h(t^\prime) + \hspace{-0.3em}\frac{\kx}{g^2}\dot{h}(t^\prime) \Big)\bar a_{\text{out}, j-1}(t^\prime )dt^\prime\\
=& \bar c_{sj-1}(T)\hspace{-0.15em}-\sin\theta \int_0^T\hspace{-0.8em} h(t^\prime+\beta)\bar a_{\text{out}, j-1}(t^\prime)dt^\prime \notag\\ \hspace{1cm}&+ O(\beta^2),
\end{align}
with $\bar c_{j0}= \cos\theta$,  $\bar a_{\text{out}, 0}= h(t)\sin\theta$, and $\beta = \kx/g^2$. The errors due to non-adiabatic corrections $(\Delta a_{\text{out}, j}, \Delta c_{s, j}(T)):=(\bar a_{\text{out} , j}(t)-a_{\text{out}, j}(t), \bar c_{sj}(T)-c_{sj}(T))$ follow the recursion relations
\begin{align}
\Delta \aout{j}(t) =& \baout{j-1}(t-2\beta)-\aout{j-1}(t) \notag\\&+ \sin\theta \Big( \bar c_{sj-1}(T)h(t-\beta ) 
-c_{sj-1}(T)h(t)\Big), \label{error1}\\
\Delta c_{sj}(T)=& \Delta c_{sj-1}(T)-\sin\theta \Big(  \int_0^T \hspace{-0.6em}h(t^\prime +\beta) \bar a_{\text{out}, j-1}(t^\prime)dt^\prime \notag\\
&-\int_0^T \hspace{-0.15em}h(t^\prime )   a_{\text{out}, j-1}(t^\prime)dt^\prime \Big).\label{error2}
\end{align}
Using notation $S_\beta  f(t) =f(t+\beta)$ and $\langle f, g \rangle:=\int f^*(x)g(x)dx$, we can simplify Eqs.~(\ref{error1}) and (\ref{error2}) as follows:
\begin{align}
\Delta \aout{j}(t) =& S_{-2\beta}\Delta \aout{j-1}(t)+(S_{-2\beta}\aout{j-1}(t)\notag \notag \\&-\aout{j-1}(t))  + \sin\theta[ \Delta c_{s, j-1}(T)S_{-\beta}h \notag \\&+ c_{s, j-1}(T)(S_{-\beta}h-h)],\\
\Delta c_{sj}(T)=& \Delta c_{sj-1}(T) - \sin\theta [\langle S_{\beta}h-h, \baout{j-1}\rangle \notag\\ &+\langle  h, \Delta\aout{j-1}\rangle].
\end{align}
We use the notation $\norm{f}_2^2:=\int \abs{f(x)}^2 dx $. Using the triangle inequality, $\langle f, g \rangle \leq \norm{f}_2\norm{g}_2$, $\norm{S_\beta f}_2 = \norm{ f}_2$, $\norm{S_\beta f - f}_2 \leq \abs{\beta} \lVert \dot f \rVert_2 + O(\beta^2)$, $\norm{\aout{j}}_2 \leq 1$, $\norm{\dot a_{\text{out}, j-1}}_2 \leq \sin(j\theta)\lVert \dot{h} \rVert_2 + O(\theta)$, and $\abs{c_{sj}(T)}\leq 1$ gives us 
\begin{align}
w_j \leq  (1+\sin\theta)w_{j-1}+\nu_j, \>\> 1 \leq j\leq 2k,
\end{align}
where $w_j= \abs{\Delta c_{sj}(T)} + \norm{\Delta \aout{j}}_2 $,  $\nu_j = 2\beta\lVert \dot h \rVert_2[\sin\theta + \sin(j\theta)]$. Using $w_0 =0$ and $\norm{\Delta a_{\text{out}, j}}_2 \leq w_j$ then gives us
\begin{align}
\norm{\Delta a_{\text{out}, 2k}}_2 \leq&   \sum_{j=1}^{2k}(1+\sin\theta)^{2k-j}\nu_j\\
=&2\beta \lVert\dot{h}\rVert_2\Big(\sin\theta \sum_{j=0}^{2k-1}(1+\sin\theta)^j \notag\\
&+\sum_{j=0}^{2k-1} \sin\left[(2k-j)\theta\right] (1+\sin\theta)^j\Big)\\
\leq & 2\beta \lVert\dot{h}\rVert_2 c_0[ 1+1/(2\sin\theta)],
\end{align}
where $c_0 = e^{\pi/2}-1$. Using $\sin\theta \geq 3/[2(2k+1)]$ for $k \geq 2$, the error due to non-adiabatic corrections can be bounded as
\begin{align}
\sqrt{1-\eta_2} \leq   c_h \frac{\kx}{g^2 T}(  10.2+5.1k), \label{worst} 
\end{align}
where $\eta_2 = \max_\tau \lvert \int_{-\infty}^\infty a_{\text{out}}(t) h^*(t+\tau) dt \rvert ^2$. We assume our mode function $h(t)$ has bandwidth $T^{-1}$, i.e., $\lVert \dot h \rVert_2 = c_h/T$. We note that this is a very loose bound and the actual errors can be many orders of magnitude smaller, e.g., for the parameters studied in Fig.~\ref{comparison}.

\section{Analysis of single-photon output in the presence of loss} \label{LossDerivation}
In this section, we perform error analysis of the single-photon generation scheme in the presence of losses. We consider spontaneous decay of the excited state $\ket{e}$ with rate $2\gamma$ and intrinsic cavity loss with rate $\kappa_{\text{i}}$. We consider the limit where the length of our Grover sequence satisfies $k \gg 1$. Here, we do not assume that the atomic state resets to $\ket{g}$ after $\ket{e}$ decays. To compute the single-photon loss to first order in the loss rates $(\kappa_{\text{i}}, \gamma)$, we use the adiabatic-limit solutions for the unitary evolution in the lossless case derived in Appendix~\ref{appbee} and compute
\begin{align}
1-\eta_3 &\approx 2\kappa_{\text{i}} \sum_{j=0}^{2k} \int_0^T \abs{c_{gj}(t)}^2 dt  + 2\gamma \sum_{j=0}^{2k}\int_0^T \abs{c_{ej}(t)}^2 dt ,
\end{align}
where $\eta_3 = \int_{-\infty}^{\infty} dt \lvert a_{\text{out}, 2k}(t) \rvert^2$,  $c_{ej}(t)$ and $c_{gj}(t)$ are the amplitudes of the states $\ket{e, 0}$ and $\ket{g,1}$, respectively, during the $j$th step of the protocol in the case of zero losses. From Appendix~\ref{appbee}, we can write the expression for $c_{gj}(t)$ to first order in $\theta$ as follows:
\begin{align}
c_{gj}(t)& \approx  h(t)\sin\theta \cos(j\theta)/\sqrt{2\kx},\>\>\> j=0, 1, \hdots, 2k.
\tag{\ref{cgjexp}}
\end{align}
This allows the simplification
\begin{align}
a_{\text{out}, j}(t) \approx  h(t)\sin\theta\sum_{m=0}^{j}\cos(m\theta), \>\> j=0, 1, \hdots, 2k.
\end{align}
We now use Eq.~(\ref{e3}) and the adiabatic limit $\kx/(g^2 T) \rightarrow 0$ to obtain
\begin{align}
c_{ej}(t)  &= \frac{i\kappa}{g}c_{gj}(t) + \frac{i\sqrt{2\kx}}{g}a_{\text{in},j}(t)\\
&=\frac{i\kappa}{g}\frac{  h(t)\sin\theta}{\sqrt{2\kx}}V_j(\theta),
\end{align}
where $a_{\text{in}, j}(t)$ is the single-photon input at the $j$th step of $U_{G_3}$.
We have
\begin{align}
V_0(\theta)&=1, \\
V_j(\theta)&= \cos (j\theta) +  \frac{2\kx}{\kappa}\sum_{m=0}^{j-1}\cos(m\theta)\\
&=  \cos(j\theta) + \frac{2\kx}{\kappa} \frac{\sin(j\theta/2)}{\sin(\theta/2)}\cos\left[(j-1)\theta/2 \right] ,
\end{align}
where $j \geq 1$. The leading term in $\theta$ for $\sum_{j=0}^{2k}\lvert V_j(\theta) \rvert^2$ is given by $ (k\kx^2)/[\kappa^2 \tan^2(\theta/2)] $. 
We then obtain
\begin{align}
1-\eta_3 &\approx  \frac{\kappa_{\text{i}}}{\kx} (k+1)\sin^2 \theta + 4kC_{\text{ex}}^{-1},
\end{align}
where $C_{\text{ex}} = g^2/(\kx\gamma)$, the first term corresponds to the intrinsic cavity loss and the second term corresponds to loss due to the excited state decay.

We now consider the case of a lossy waveguide, where the transmission probability per trip is $1-\epsilon_t$. The total single-photon loss through the waveguide for our protocol, $L_\mathrm{w}$, with $\epsilon_t k \ll 1$ and $k \gg 1$, can be approximated as
\begin{align}
L_{\text{w}} & \approx \epsilon_t \sum_{j=0}^{2k}\int_0^T \lvert a_{\text{out}, j}(t) \rvert^2 dt\\
&= \epsilon_t [k + 3/2 + O(1/k)].
\end{align}

\section{Initial state preparation and error analysis for the NOON state}\label{errorNOON}
In this section, we bound the error in the state fidelity $F_{\text{NOON}} =|\langle \psi_{\text{NOON}}| U_n^{(2)} U_n^{(1)} | \psi_0 \rangle|^2$, where $\ket{\psi_{\text{NOON}}} \propto \ket{0,n}_{1,2} + \ket{n,0}_{1,2}$ is the NOON state, and $\ket{\psi_0} \propto \ket{0, \lambda}_{1,2}+\ket{\lambda, 0}_{1,2}$, with the two cavity modes labeled by subscripts 1 and 2.  Here,  $U_n^{(2)}U_n^{(1)}$ is the protocol described in Sec.~\ref{res1}.

We first compute an upper bound for $  1- \lvert \bra{0}U_n^{(i)}\ket{0} \rvert $. From Eq.~(\ref{G2fp}), we have 
\begin{align}
U_n^{(i)}\ket{0}_i = (-1)^l \prod_{j=1}^{l} Q_j \ket{0}_i,
\end{align}
where $Q_j = R_i(\alpha_j)R_1(\beta_j)$. We then define $A_t:=\bra{0}Q_t \hdots Q_1 \ket{0}$ with $A_0 := \bra{0}I \ket{0}=1$. We then have
\begin{align}
 \abs{A_l-1}   &\leq \sum_{t=1}^{l}\abs{A_t - A_{t-1}} \label{Albm1}.
\end{align}
We now note that $A_{t}-A_{t-1} = \bra{0}(Q_t - I)(Q_{t-1}\hdots Q_1)\ket{0}_i$, and using the Cauchy-Schwarz inequality gives us $\abs{A_t - A_{t-1}} \leq \lVert (Q_t^\dagger - I) \ket{0}_i \rVert_2$. We first note that
\begin{align}
(Q_j^\dagger - I)\ket{0}_i &= [R_1(-\beta_j) R_i(-\alpha_j) - I]\ket{0}_i\\
&=[R_1(-\beta_j) R_i(-\alpha_j) -R_1(-\beta_j)]\ket{0}_i\\
&=R_1(-\beta_j)(R_i(-\alpha_j) - I)\ket{0}_i\\
&= -(1-e^{-i\alpha_j})R_1(-\beta_j)\braket{\lambda}{0}_i \ket{\lambda}_i,
\end{align}
where $R_i$ is given in Eq.~(\ref{rideffock}), $\ket{\lambda}$ is a coherent state with amplitude $\lambda = \sqrt{n}$, and we use the fact that $R_1(-\beta) = e^{ -i\beta \ketbra{n}{n}}$ acts as the identity on $\ket{0}_i$. We then have $\lVert (Q_j^\dagger - I)\ket{0}_i \rVert_2 \leq 2 \abs{\braket{\lambda}{0}_i} \leq 2 e^{-n/2}$. We can then bound $\abs{A_l - 1}$ from Eq.~(\ref{Albm1}) as follows:
\begin{align}
\abs{A_l - 1} \leq \epsilon,\\
\epsilon = 2le^{-n/2}. \label{Albound} 
\end{align}
Using $1-\abs{z} \leq \abs{1-z}$ gives us $1-\lvert\bra{0}U_{n}^{(i)} \ket{0} \rvert  \leq \epsilon$.

We now compute a lower bound on the state fidelity of our prepared state given by $F_{\text{NOON}} = \lvert\bra{\psi_{\text{NOON}}}U_n^{(2)}U_n^{(1)}\ket{\psi_0} \rvert^2$. First note that we can compute $\bra{\psi_{\text{NOON}}}U_n^{(2)}U_n^{(1)}\ket{\psi_0}$ as follows:
\begin{align}
&\bra{\psi_{\text{NOON}}}U_n^{(2)}U_n^{(1)}\ket{\psi_0} \notag
\\=&\frac{1}{2\bar N} \big( \bra{0, n}U_{n}^{(2)}U_{n}^{(1)}\ket{0, \lambda} + \bra{n, 0}U_{n}^{(2)}U_{n}^{(1)}\ket{0, \lambda} \notag \\
&+ \bra{0, n}U_{n}^{(2)}U_{n}^{(1)}\ket{\lambda, 0}+ \bra{n, 0}U_{n}^{(2)}U_{n}^{(1)}\ket{\lambda, 0} \big) \label{combinemode}\\
=& \frac{1}{\bar N}\big( \bra{0}U_n^{(1)} \ket{0} \bra{n}U_n^{(2)} \ket{\lambda} + \bra{n}U_n^{(1)} \ket{0}\bra{0}U_n^{(2)}\ket{\lambda}\big) \label{nosuper},
\end{align}
where $\bar{N} = \sqrt{1+e^{-n}}$. In Eq.~(\ref{combinemode}), we used the fact that the unitaries $U_n^{(i)}$ are identical for all modes $i$. We then use the triangle inequality $\abs{z_1+z_2} \geq \abs{z_1}-\abs{z_2}$ and the bounds $\lvert\bra{0}U_n^{(i)}\ket{0}_i \rvert \geq 1-\epsilon$ from Eq.~(\ref{Albound}) (using $1-\abs{z} \leq \abs{1-z}$), $\lvert \bra{n}U_n^{(i)}\ket{\lambda}_i \rvert \geq \sqrt{1-\delta^2}$ from Eq.~(\ref{lbound}), $\lvert \bra{n}U_n^{(i)}\ket{0}_i \rvert^2 \leq  1-\lvert \bra{0}U_n^{(i)}\ket{0}_i \rvert^2  \leq  2\epsilon$ from Eq.~(\ref{Albound}), and $\lvert\bra{0}U_n^{(2)}\ket{\lambda} \rvert^2 \leq 1-\lvert \bra{n}U_n^{(2)}\ket{\lambda}  \rvert^2 \leq \delta^2$ to obtain
\begin{align}
\lvert \bra{\psi_{\text{NOON}}}U_n^{(2)}U_n^{(1)}\ket{\psi_0} \rvert \geq& [(1-\epsilon)(1-\delta^2)-\sqrt{2\epsilon}\delta] \notag\\ &\times(1-e^{-n}/2).
\end{align}
The above result gives us the following lower bound on the fidelity
\begin{align}
\lvert \bra{\psi_{\text{NOON}}}U_n^{(2)}U_n^{(1)}\ket{\psi_0} \rvert^2 \geq  1-2[y +e^{-n}/2], \label{expFidNoon}
\end{align}
where $y = \delta^2 + \epsilon +\sqrt{2\epsilon}\delta - \epsilon \delta^2$. Here, $\delta$ is related to $l$ via Eq.~(\ref{2ldrel}). In the limit $n \gg 1$, we ignore terms that scale as $e^{-n/2}$ and approximate $y(\epsilon, \delta)$ in terms of $n$ and $l$ as follows:
\begin{align}
y \approx 4\exp[-2l/(cn^{1/4})]+4e^{-n/4} \exp[-l/(cn^{1/4})]\sqrt{l},
\end{align}
where $c = (2\pi)^{1/4}/2$ and $\delta \approx 2\exp[-l/(cn^{1/4})] $ for $n \gg 1$.
We can differentiate the above with respect to $l$ and find that the optimal scaling of $l$ that gives $y \sim e^{-n/2}$ (since we ignored terms that scale as $e^{-n/2}$) is given by $l=cn^{5/4}/4$. The error with the optimal $l(n)$ for $n\gg1$ is given by $1-F_{\text{NOON}} \lesssim  e^{-n/2}n^{5/8}(4\sqrt{c}+n^{5/8}c) $.
\end{document}